\documentclass[twocolumn,showpacs,preprintnumbers,amsmath,amssymb,prb]{revtex4-1}
\usepackage{amsmath}
\usepackage{natbib}
\usepackage{amssymb}
\usepackage[dvips]{color}
\usepackage{graphicx}
\usepackage{dcolumn} 
\usepackage{bm} 
\usepackage{hyperref}
\usepackage{longtable}

\begin{document}
\texttt{}

\title{Separating the bulk and surface \textit{n}- to \textit{p}-type transition in the topological insulator GeBi$_{4-x}$Sb$_x$Te$_7$}

\author{Stefan Muff$^{1,2}$}
\author{Fabian von Rohr$^{1,3}$}
\author{Gabriel Landolt$^{1,2}$}
\author{Bartosz Slomski$^{1,2}$}
\author{Andreas Schilling$^{1}$}
\author{Robert J. Cava$^{3}$}
\author{J\"urg Osterwalder$^{1}$}
\author{J. Hugo Dil$^{1,2}$}

\affiliation{
$^{1}$Physik-Institut, Universit\"at Z\"urich, Winterthurerstrasse 190, 
CH-8057 Z\"urich, Switzerland 
\\ 
$^{2}$ Swiss Light Source, Paul Scherrer Institut, CH-5232 Villigen, 
Switzerland
\\
$^{3}$ Department of Chemistry, Princeton University, Princeton, New Jersey 08544, USA 
}

\date{\today}

\begin{abstract}

We identify the multi--layered compound GeBi$_4$Te$_7$ to be a topological insulator with a freestanding Dirac point, slightly above the valence band maximum, using angle--resolved photoemission spectroscopy (ARPES) measurements. The spin polarization satisfies the time reversal symmetry of the surface states, visible in spin--resolved ARPES. 
For increasing Sb content in GeBi$_{4-x}$Sb$_x$Te$_7$ we observe a transition from \textit{n}- to \textit{p}-type in bulk sensitive Seebeck coefficient measurements at a doping of $x=0.6$. In surface sensitive ARPES measurements a rigid band shift is observed with Sb doping, accompanied by a movement of the Dirac point towards the Fermi level. Between $x=0.8$ and $x=1$ the Fermi level crosses the band gap, changing the surface transport regime.
This difference of the \textit{n}- to \textit{p}-type transition between the surface region and the bulk is caused by band bending effects which are also responsible for a non-coexistence of insulating phases in the bulk and in the near surface region.

\end{abstract}

\pacs{72.15.Jf, 79.60.Bm, 73.20.At}

\maketitle

\section{Introduction}

Since their theoretical prediction and experimental discovery, three--dimensional topological insulators have been established as a new field in solid state physics. In topological insulators a strong spin-orbit coupling is responsible for spin--polarized surface states with a Dirac dispersion in the bulk band gap, forming a conductive surface.\cite{Kane:2005a, Hasan:2010, Qi:2011} 
For experimental devices investigating the transport properties of this kind of material, samples with the Fermi level inside the band gap are of special interest to access the surface states directly. The method of choice to achieve this condition is to tune the Fermi level of a known topological insulator by doping.\cite{Zhang:2011b, Souma:2012, Arakane:2012} Currently the most popular measurement method in these experiments is angle-resolved photoemission spectroscopy (ARPES), which provides a picture of the electron energy distribution in \textit{k}-space in the near surface region of the material.\cite{Hsieh:2008, Hsieh:2009, Xia:2009, Chen:2009, Nishide:2010, Chen:2010b} The information obtained from these experiments is generally transfered to the electron dispersion in the bulk, but surface-related effects can be responsible for a remarkable difference between the electronic structure in the bulk and in the surface region.

The current work focuses on the properties of the topological insulator GeBi$_4$Te$_7$.\cite{Kuznetsova:2000,Zhang:2009,Xu:2011arXiv} We identify this material as an ideal starting point to shift the Dirac point towards the Fermi level, due to its electronic structure with the freestanding Dirac point at the $\bar{\Gamma}$-point. In spin--resolved ARPES (SARPES) measurements of the surface states we distinguish a helical orientation of the in--plane spin polarization and a buckling in the out--of--plane direction.
Our data tracks the band structure evolution of GeBi$_{4-x}$Sb$_x$Te$_7$ under Sb doping with surface sensitive ARPES measurements and bulk sensitive Seebeck coefficient measurements. In this compound the Sb atoms replace Bi atoms, leading to a defect--induced transport regime transition without a change of crystalline structure but a slight compression of the unit cell.\cite{vRohr:2012} We are able to distinguish the \textit{n}- to \textit{p}- type transition at $x=0.6$ in the bulk and at $x=0.95$ at the surface. The combined interpretation of the results from the two measurement methods shows a mismatch of insulating phases in the bulk and in the surface region. This lack of a simultaneous insulating surface region and insulating bulk is caused by band bending at the surface and is a serious restriction in the search of tunable topological insulators with the Fermi energy located in the band gap.\cite{Hsieh:2009N}

\section{Experiments}

The here presented ARPES measurements were performed at the Surface and Interface Spectroscopy beamline of the Swiss Light Source using the COPHEE end station with \textit{p}-polarized synchrotron light of 20~eV.\cite{Hoesch:2002} The Omicron electron energy analyzer of this end station has an energy resolution of 25~meV and an acceptance angle of 0.5$^\circ$. A movement of the sample with respect to six axes is possible with a CARVING manipulator holding the sample temperature at 23 K. Two classical 40kV Mott detectors mounted orthogonally to each other give the possibility to determine the spin polarization of the photoelectrons with respect to all three spatial directions. In the SARPES mode the energy resolution of the analyzer is 60~meV with an acceptance angle of 1.5$^\circ$.

\begin{figure*}[htbp]
	\centering
	\includegraphics[width=0.98\textwidth]{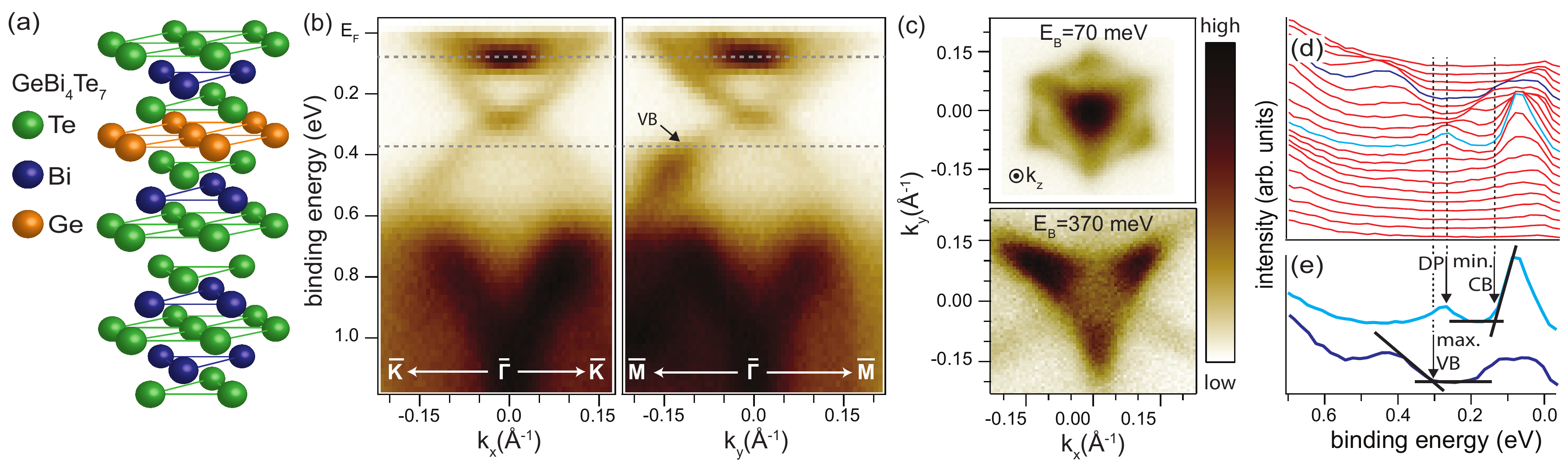}
\caption{(color online) (a) Crystal structure of GeBi$_4$Te$_7$. (b) Band dispersion along the high symmetry directions $\bar{\Gamma}\bar{K}$ and $\bar{\Gamma}\bar{M}$. (c) Constant energy cuts in the CB region at $E_b=70~$meV and at $E_b=370~$meV in the VB region. (d) Energy distribution curves in $\bar{\Gamma}\bar{M}$ direction. (e) Energy distribution curves at normal emission (bright blue) and at $k=-0.11$~\AA$^{-1}$ (dark blue) in $\bar{\Gamma}\bar{M}$ direction with marked points of interest.} 
\label{fig1}	
\end{figure*}

The samples were synthesized by heating stoichiometric mixtures of Ge (99.99\% pure), Te (99.99\%), Sb (99.999\%) and Bi (99.999\%) to 950$~^\circ$C for 12~h in 12~mm diameter quartz tubes under vacuum. The crystal growth took place via slow cooling from 950$~^\circ$C to 450$~^\circ$C. The best quality crystals were obtained with a cooling rate of 50$~^\circ$C/h, annealing at 450$~^\circ$C for one week and then quenching in cold water.\cite{vRohr:2012} Slower cooling rates yielded crystals of lower quality. Annealing and a homogenous heat distribution in the furnace were found to be crucial for obtaining crystals of reproducible high quality. Temperature dependent measurements of the transport properties were carried out in a Quantum Design physical property measurement system (PPMS) equipped with the thermal transport option (TTO).

For ARPES measurements the samples were glued on copper sample holders and cleaved under ultra high vacuum conditions at low temperatures to produce clean surfaces. GeBi$_4$Te$_7$ is a layered material consisting of a quintuple layer of Bi$_2$Te$_3$ and a septuple layer of GeBi$_2$Te$_4$(see Fig. \ref{fig1}(a)), isostructural to PbBi$_4$Te$_7$.\cite{Eremeev:2012} This stack of 12 layers results in a large unit cell with $c=23.89~$\AA$~$ and $a=4.29~$\AA.\cite{vRohr:2012,Shelimova:2004,Kuznetsov:1998} 
Under Sb doping to GeBi$_{4-x}$Sb$_x$Te$_7$, the dopant Sb is expected to randomly substitute Bi. 
Cleaving takes place in the weakly bonding van der Waals gap between the quintuple and septuple layers. Due to the self--protecting behavior of the topological state the exact kind of termination is not important for this study. Because they are energetically similar, a mixture of both terminations is expected on the surface of the cleaved sample.\cite{Eremeev:2012}

\section{Results and Discussion}

Figs. \ref{fig1}(b) and (c) show ARPES results of the parent compound GeBi$_4$Te$_7$. 
Band dispersions measured along the high symmetry lines $\bar{\Gamma}\bar{M}$ and $\bar{\Gamma}\bar{K}$ (Fig.\ref{fig1}(b)) show the surface bands forming the typical X-shaped Dirac cones centered at $\overline{\Gamma}$ within the bulk band gap. 
In the band map along $\bar{\Gamma}\bar{M}$ it becomes clear that the Dirac point is only slightly above the valence band (VB) maximum which is visible on the left side (see mark in Fig. \ref{fig1}(b)) at the used photon energy of $h\nu=20~$eV (see also next paragraph).
The constant energy map at $E_b=70~$meV in Fig. \ref{fig1}(c) consists of a weak hexagonal contour of the surface states which enclose the conduction band (CB). The CB form a triangular structure with edges pointing in the $\bar{\Gamma}\bar{M}$ direction. 
In the constant energy cut at $E_b=370~$meV in Fig. \ref{fig1}(c) we see the VBs forming the dominant triangular structure, pointing in the $\bar{\Gamma}\bar{M}$ direction and a weak triangle rotated by 180$^\circ$. The VBs overlap the surface bands which are, due to the high intensity of the VB, not visible at this binding energy.

The energy distribution curves in the $\bar{\Gamma}\bar{M}$ direction plotted in Fig.\ref{fig1}(d) emphasize that below the Dirac point the surface states have low intensity, but they are nicely pronounced from the Dirac point heading towards the Fermi energy. 
The energy distribution curves at normal emission and at $k_{||}=-0.11$~\AA$^{-1}$ in $\bar{\Gamma}\bar{M}$ direction, where the maximum of the VB is located, show that at this photon energy the edges of the bulk bands are situated at $E_b=(300\pm25)~$meV for the VB and $E_b=(130\pm25)$~meV for the CB (Fig. \ref{fig1}(e)). This means the measured gap is in the range of $E_g=(170\pm50)~$meV with the Dirac point situated at $E_b=(280\pm25)$~meV. The measured band structure, especially the VB visible in the $\bar{\Gamma}\bar{M}$ direction, is similar to the band structure of the chemically and structurally related GeBi$_2$Te$_4$, but the energy positions of the CB minimum, the VB maximum as well as the Dirac point differ.\cite{Okamoto:2012}

\begin{figure}[htbp]
	\centering
		\includegraphics[width=0.48\textwidth]{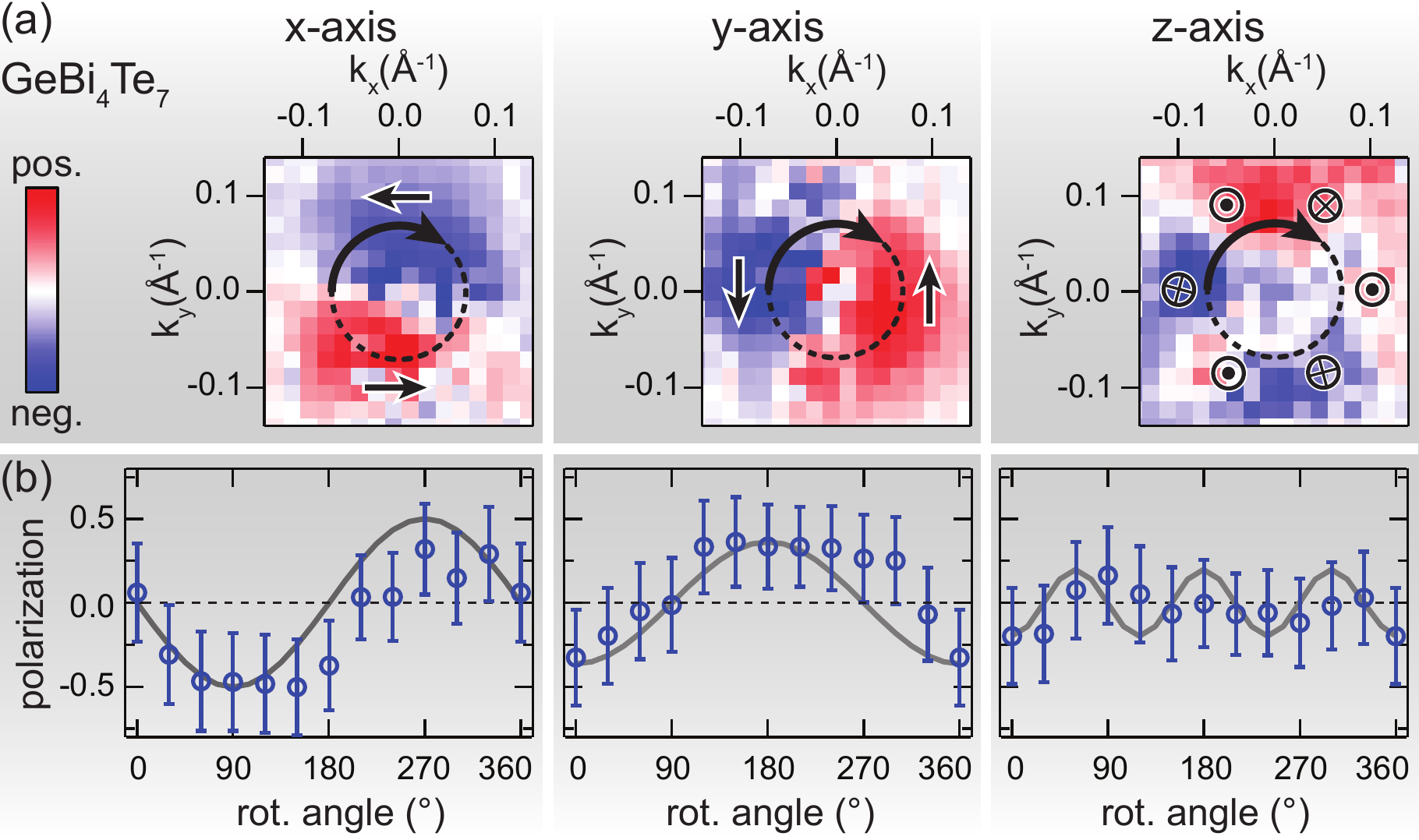}
	\caption{(color online) (a) Spin polarization maps in the $k_x$-$k_y$ plane at $E_b=200~$meV of GeBi$_4$Te$_7$ for the three spatial components weighted by the total intensity. (b) Polarization along the circle marked in (a) with a radius of $0.07$~\AA$^{-1}$ in the direction of the arrow. The sinusoidal grey lines are guides to the eye.}
	\label{fig2}
\end{figure}

SARPES measurements at a constant binding energy of 200~meV, between the Dirac point and the CB minimum, are displayed in Figs. \ref{fig2}(a) and (b). Only a low contribution of the bulk states to the spectral weight is expected at this energy. Accordingly, the measured spin polarization can be attributed to the surface states. The spin polarization is displayed based on a right--handed coordinate system with respect to the sample normal. 
The first two graphs in Fig.\ref{fig2}(a) show the spin polarization map of the photoelectrons in the in--plane directions x and y. In the third one the out--of--plane polarization map along the z-direction is displayed. The spin polarization along a circle with a radius of $0.07$~\AA$^{-1}$ in clockwise direction are displayed in Fig.\ref{fig2}(b). This circle corresponds roughly to the hexagonal shape of the surface bands at this binding energy.

The in--plane polarization shows a helical spin orientation. This results in a splitting of the circle in the x- and y-polarization maps into semicircles with positive and negative polarization and a sinusoidal behavior with a periodicity of 360$^\circ$ in the x- and y-polarization along the circle in Fig.\ref{fig2}(b). 
Along the high symmetry directions the in--plane components of the spin are completely polarized in the x- (along $\bar{\Gamma}\bar{M}$) and y-direction (along $\bar{\Gamma}\bar{K}$), respectively, indicated with the arrows in the corresponding polarization maps (see Fig.\ref{fig2}(a)). An effect of the \textit{p}-polarized light on the spin polarization of the photoelectrons resulting in radial in--plane polarizations between the high symmetry directions as theoretically suggested by \citet{Park:2012a} was not observed. In the polarization map of the out--of--plane direction z, a sixfold symmetric buckling of the spin is visible (see right panel in Fig.\ref{fig2}(a)). This up and down of the spin in z-direction is a result of the warping of the hexagon--shaped constant energy surface of the surface states.\cite{Fu:2009w, Souma:2011, Eremeev:2012, Dil:2009R} 

\begin{figure*}[htbp]
\centering
\includegraphics[width=0.98\textwidth]{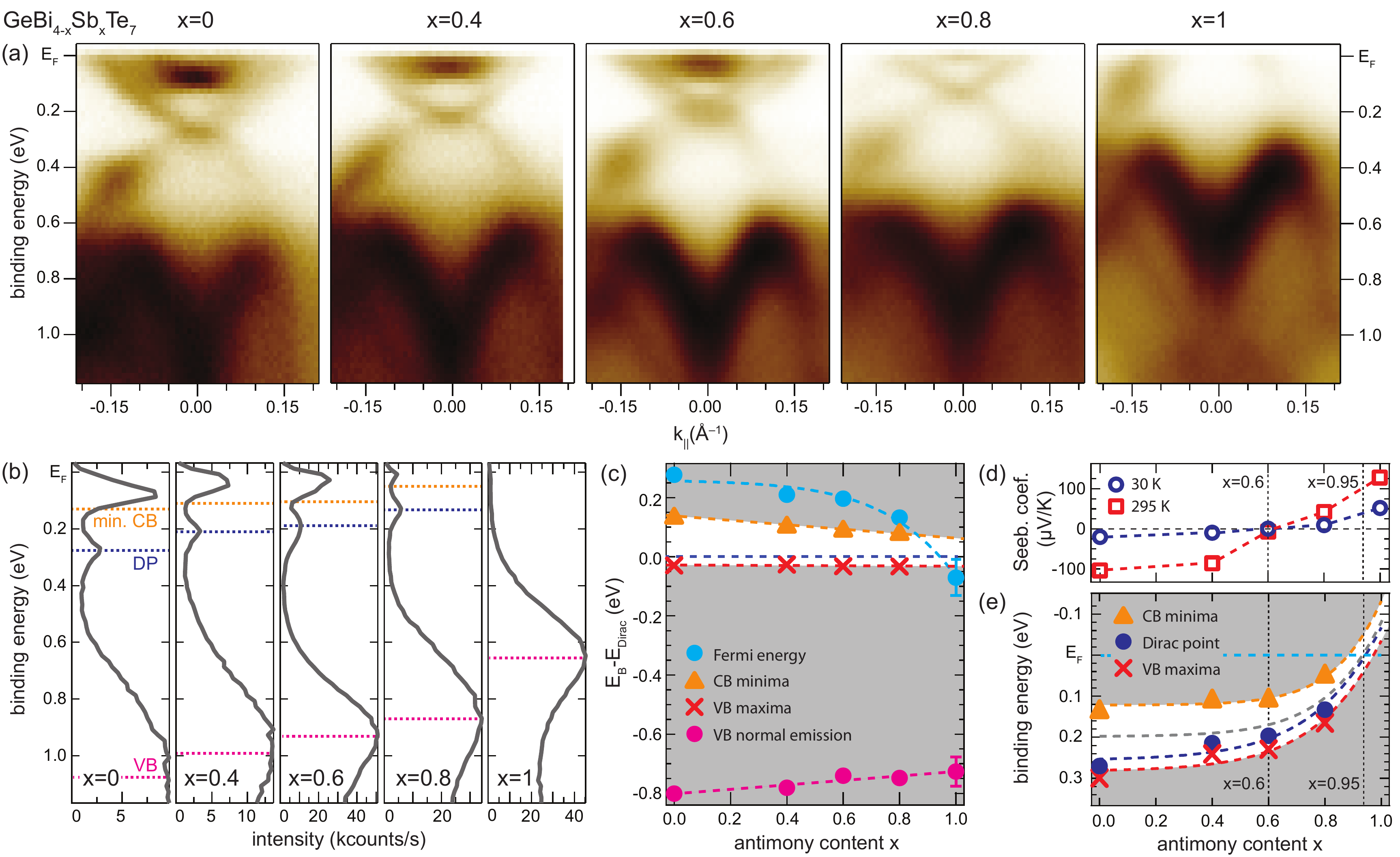}
\caption{(color online) (a) Band dispersions around normal emission for five different compounds of GeBi$_{4-x}$Sb$_x$Te$_7$ with $x=$0, 0.4, 0.6, 0.8 and 1. (b) Energy distribution curves at $k=0~$\AA$^{-1}$ for the different Sb contents with marked points of interest: min of CB, Dirac point and VB feature.
(c) Energetic position of the Fermi levels with respect to the Dirac point fitted with a sigmoid function. The energetic position of the CB minima, the VB maxima and the VB at normal emission with respect to the Dirac point fitted with linear functions. The VB and CB region are shaded grey. Errors are of the size of the marks ($\pm~25meV$) or indicated with error bars. 
(d) Composition dependence of the Seebeck coefficient of the measured samples at 30 K and 295 K. (e) Shift of the binding energy of the CB minimum, the Dirac point, the VB maximum and corresponding fitted sigmoid curves (dashed) with respect to the Sb content $x$. The dashed grey line marks the center of the band gap. The VB and CB region are shaded grey. Errors are of the size of the marks ($\pm~25meV$).} 
\label{fig3}
\end{figure*}

Next we focus on the effects of Sb doping on the band structure of GeBi$_{4-x}$Sb$_x$Te$_7$. The following ARPES data provide evidence that under this doping it is possible to alter the band structure and tune the Fermi level towards the Dirac point. Fig.\ref{fig3}(a) shows the ARPES measurements of five different compounds starting with the undoped sample GeBi$_4$Te$_7$ up to GeBi$_3$SbTe$_7$ ($x=1$). Comparing the first two data sets, the undoped sample and the lightly doped sample ($x=0.4$), the Dirac point as well as the position of the bulk VB shifted slightly towards the Fermi level. This rigid band shift continues at a doping of $x=0.6$ and at $x=0.8$ the CB is almost completely shifted above the Fermi level. Finally at the largest doping, in GeBi$_3$SbTe$_7$, the Dirac point is no longer occupied and only the lower part of the Dirac cone and the maxima of the VB cross the Fermi level. At a doping between $x=0.8$ and $x=1$ the Dirac point crosses the Fermi level. 

For an analysis of the band shift under Sb doping, the energetic position of Dirac point (x=0, 0.4, 0.6 and 0.8) and the broad VB feature at normal emission (all x) are fitted with Gaussian functions in the energy distribution curves in Fig. \ref{fig3}(b). The position of the CB minima at normal emission an the VB maxima at $k_{||}=-0.11$~\AA$^{-1}$ were located the same way as shown in Fig.\ref{fig1}(e) for x=0, 0.4, 0.6 and 0.8. 
The relative energetic positions of these points in the different band structures are plotted with respect to the Sb content in Fig.\ref{fig3}(c), using the Dirac point as the reference. 
In Fig.\ref{fig3}(c) we see, that the position of the broad VB feature at normal emission with respect to the Dirac point is a linear function of the Sb content. 
With an extrapolation of this linear function to $x=1$ the position of the VB at normal emission and accordingly the position of the Dirac point located at $E_b=(-70\pm60)$~meV above the Fermi level can be distinguished for this compound. Comparing this position with the energetic position of the Dirac point at $E_b=(280\pm25)$~meV for GeBi$_{4}$Te$_7$, a band structure shift of $(350\pm65)$~meV over the whole doping range is observed. 
The energetic position of the VB maxima and the CB minima relative to the Dirac point change linearly with the Sb content as well (see Fig.\ref{fig3}(c)). By the help of fitted linear functions the energetic positions of the VB maxima and the CB minima for GeBi$_3$SbTe$_7$ can be extrapolated (see Fig.\ref{fig3}(c)). The binding energy of the Dirac point as a function of the Sb content is fitted with a sigmoid function providing a Fermi level crossing at a doping level of $x=0.95\pm0.03$ (see Fig.\ref{fig3}(e)), accompanied by a \textit{n}- to \textit{p}-type transition of the surface transport regime.

This \textit{n}- to \textit{p}-type crossover is also observed in the bulk transport properties. In Fig.\ref{fig3}(d) the composition dependence of the of GeBi$_{4-x}$Sb$_x$Te$_7$ solid solution is shown. We observe a sign change and consequently a change of the transport regime at a doping of $x=0.6$ at 30 K where the ARPES measurements were done, as well as at room temperature (see Fig.\ref{fig3}(d)). This discrepancy between the \textit{n}- to \textit{p}-type transition measured at the surface and in the bulk can be explained by band bending at the sample surface, commonly known in semiconductor physics as an effect of the surface states or defect induced as observed for Bi$_2$Se$_3$.\cite{Zhang:2012, Bianchi:2010, King:2011}. At a doping of $x=0.6$, where the bulk \textit{n}- to \textit{p}-type transition takes place we expect the band gap center of the bulk at the Fermi level. By the help of the position of the band gap center measured in the surface region for a doping of $x=0.6$ (see Fig.\ref{fig3}(e), dashed grey line) the downwards bending of the bands at the surface can be estimated to be 170~meV (difference between the band gap center in the surface region and in the bulk). 

The measured band gap marked in Figs. \ref{fig3}(c) and (e) by the white region decreases linearly with Sb doping. The band gap of the undoped sample of $E_{g}=(170\pm35)$~meV is narrowing to $(95\pm70)$~meV for GeBi$_3$SbTe$_7$ if the positions of the CB minima and VB maxima are extrapolated. This narrowing of the band gap is explained by the lower atomic number of Sb compared to Bi leading to a weaker spin--orbit coupling which is responsible for the energy splitting of the bands enclosing the gap. 
Based on this narrowing we expect to observe the Fermi level inside the band gap in ARPES measurements for a doping between $x=0.87\pm0.02$ and $x=0.98\pm0.04$. The observed band gap of the undoped sample is of the order of the band bending at the surface and for doped samples between $x=0.6$ and $x=1$ the band gap is even smaller than the band bending, assuming a constant band bending for all doping levels. Accordingly there can be no doping where a insulating bulk occurs simultaneously with a insulating surface region.

\begin{figure}[htbp]
	\centering
		\includegraphics[width=0.48\textwidth]{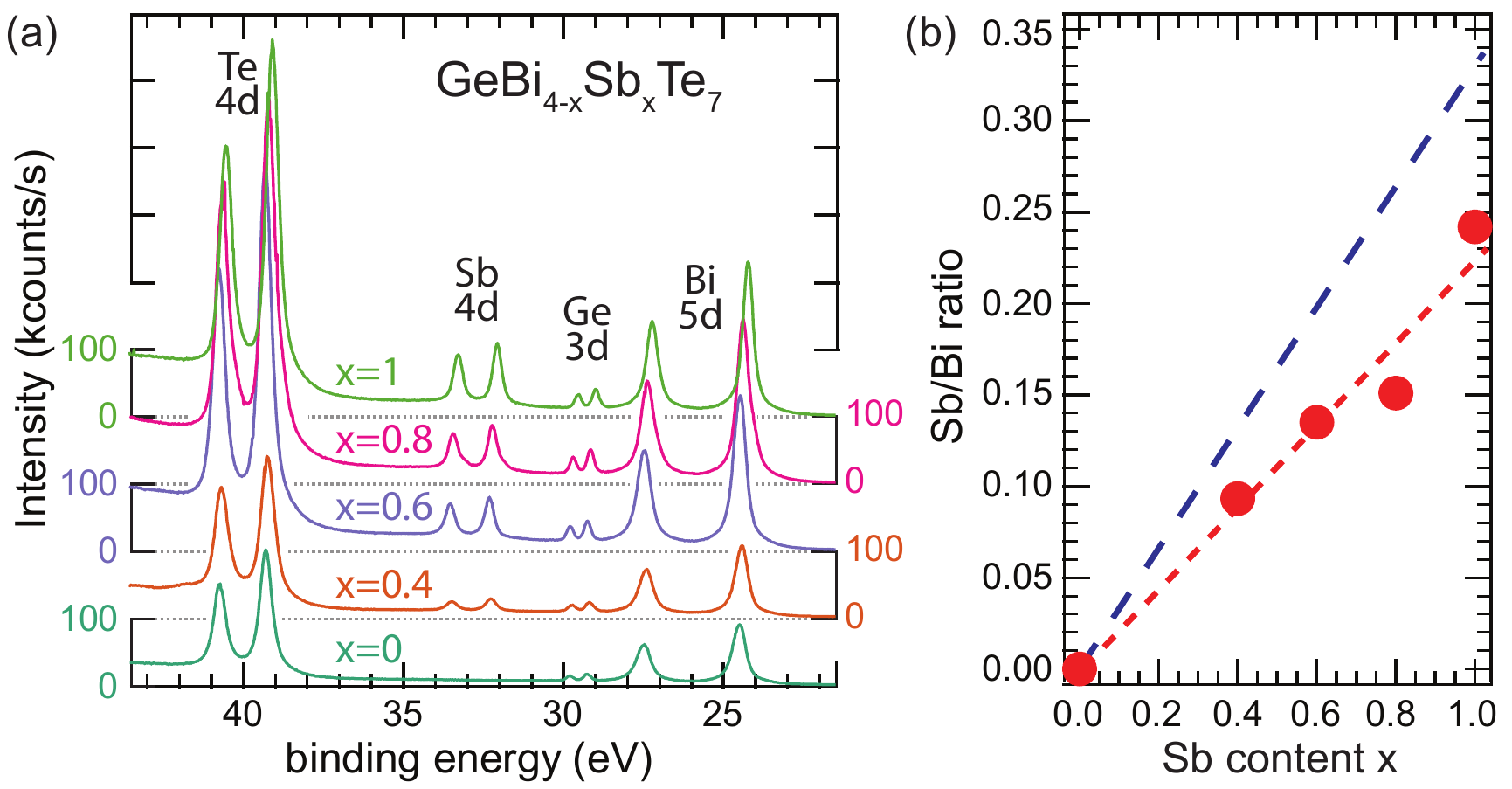}
	\caption{(color online) (a) XPS spectra in the region of interest for the different compounds measured with $h\nu=80~$eV. (b) Sb/Bi ratio determined from the measurements, the dashed red line is a straight line fitted to the data. The dashed blue line shows the stoichiometric ratio of the samples.}
	\label{fig4}
\end{figure}

To gain further understanding of the mechanism responsible for the band shift under Sb doping the compositions of the samples were investigated by X--ray photoelectron spectroscopy (XPS). XPS data in the energy region of the Te 4d, Sb 4d, Ge 3d and Bi 5d core levels measured with a photon energy of $h\nu=80~$eV are shown in Fig.\ref{fig4}(a). In first order approximation we assume that the Sb and Bi atoms are distributed in the same atomic layers. Under these conditions the ratio of Sb to Bi is given by the peak intensities of the Sb 4d and Bi 5d core levels weighted by the respective cross sections (see Fig.\ref{fig4}(b)). The measured ratios show a linear behavior, although their value is lower than expected from the stoichiometric composition. 
In XPS measurements the inelastic scattering length of the emitted electrons depends on the position of the emitting atom in the crystal. Because the exact positions of the Sb and Bi atoms in the crystal with respect to the surface and to each other is unknown, the modulation of the XPS intensities due to inelastic scattering is not known precisely which may explain the difference between the measured and the stoichiometric ratio.

The dopant Sb has the same valency as Bi, and thus it acts neither as an acceptor nor as a donor compared to Bi in GeBi$_{4-x}$Sb$_x$Te$_7$. Therefore the change in band filling as a function of increasing Sb content is not caused by carrier doping but rather by the type of defects caused by the Bi/Sb replacement. Calculations of the formation energies of the prototype case of (Bi$_{1-x}$Sb$_x$)$_2$Te$_3$ by \citet{Niu:2012} show that in the undoped Bi$_2$Te$_3$ sample Te vacancies and Te$_{\text{Bi}}$ antisite defects (Bi replaced by Te) are dominant and are supporting the $n$-type conductance of the compound. With increasing Sb content these defects are energetically less favorable but the building of Sb$_{\text{Te}}$ antisite defects is preferred. With less Te$_{\text{Bi}}$ and more Sb$_{\text{Te}}$ antisite defects the system becomes $p$-type. Due to the similarities of this system with GeBi$_{4-x}$Sb$_x$Te$_7$ we expect the same type of defects to be responsible for the observed $n$- to $p$-type transition.\cite{Niu:2012, Huang:2012b, Zhang:2011b} The additional role of Ge related antisite defects could be the aim of further theoretical investigations.

The calculation of \citet{Niu:2012} demonstrates also that a small change in the sample growth procedure, providing more energy during the sample formation, can open the possibility that other defects with higher formation energy may form. This as well as the sensitivity of the system to small stoichiometric changes \cite{Kuznetsov:1998} is why a highly controlled sample preparation procedure with identical parameters for all the different doping levels is mandatory to reproducibly observe this change in electronic structure. We also studied samples of poorer quality who show overall features similar to what we report here, but the smooth transition of the position of the Dirac point through the Fermi energy as a function of Sb content is obscured by sample inhomogeneity.
\newline
~\\
\section{Conclusion}

We present (S)ARPES data of the multi--layered material GeBi$_4$Te$_7$ and a study of the change of the band structure under doping with Sb to GeBi$_{4-x}$Sb$_x$Te$_7$. Our data allocates spin--polarized surface states forming a freestanding Dirac point in the band gap of GeBi$_4$Te$_7$.
Upon Sb doping we find a rigid shift of the band structure and the Dirac point crossing the Fermi level between a doping of $x=0.8$ and $x=1$ giving the possibility to change the surface transport regime from \textit{n}- to \textit{p}-type by doping. Responsible for the observed band shift is the introduction of different types of defects in the crystal structure under Sb doping. The \textit{n}- to \textit{p}-type transition observed in surface sensitive ARPES is clearly distinct from the transition in the bulk, visible in Seebeck coefficient measurements at $x=0.6$. Combining these two measurements we are able to distinguish a band bending of 170~meV at the sample surface. The concluded non--coexistence of a insulating phase in the near surface region and in the bulk demonstrates the importance of the combination of bulk and surface sensitive measurements to get materials with a Fermi level in the band gap to access the topological surface states directly in transport measurements.

\section{Acknowledgment}

We thank C. Hess, F. Dubi and M. Kropf for the technical support. This work was supported by the Swiss National Science Foundation.

\bibliographystyle{apsrev4-1}


\begin{thebibliography}{31}%
\makeatletter
\providecommand \@ifxundefined [1]{%
 \@ifx{#1\undefined}
}%
\providecommand \@ifnum [1]{%
 \ifnum #1\expandafter \@firstoftwo
 \else \expandafter \@secondoftwo
 \fi
}%
\providecommand \@ifx [1]{%
 \ifx #1\expandafter \@firstoftwo
 \else \expandafter \@secondoftwo
 \fi
}%
\providecommand \natexlab [1]{#1}%
\providecommand \enquote  [1]{``#1''}%
\providecommand \bibnamefont  [1]{#1}%
\providecommand \bibfnamefont [1]{#1}%
\providecommand \citenamefont [1]{#1}%
\providecommand \href@noop [0]{\@secondoftwo}%
\providecommand \href [0]{\begingroup \@sanitize@url \@href}%
\providecommand \@href[1]{\@@startlink{#1}\@@href}%
\providecommand \@@href[1]{\endgroup#1\@@endlink}%
\providecommand \@sanitize@url [0]{\catcode `\\12\catcode `\$12\catcode
  `\&12\catcode `\#12\catcode `\^12\catcode `\_12\catcode `\%12\relax}%
\providecommand \@@startlink[1]{}%
\providecommand \@@endlink[0]{}%
\providecommand \url  [0]{\begingroup\@sanitize@url \@url }%
\providecommand \@url [1]{\endgroup\@href {#1}{\urlprefix }}%
\providecommand \urlprefix  [0]{URL }%
\providecommand \Eprint [0]{\href }%
\providecommand \doibase [0]{http://dx.doi.org/}%
\providecommand \selectlanguage [0]{\@gobble}%
\providecommand \bibinfo  [0]{\@secondoftwo}%
\providecommand \bibfield  [0]{\@secondoftwo}%
\providecommand \translation [1]{[#1]}%
\providecommand \BibitemOpen [0]{}%
\providecommand \bibitemStop [0]{}%
\providecommand \bibitemNoStop [0]{.\EOS\space}%
\providecommand \EOS [0]{\spacefactor3000\relax}%
\providecommand \BibitemShut  [1]{\csname bibitem#1\endcsname}%
\let\auto@bib@innerbib\@empty
\bibitem [{\citenamefont {Kane}\ and\ \citenamefont {Mele}(2005)}]{Kane:2005a}%
  \BibitemOpen
  \bibfield  {author} {\bibinfo {author} {\bibfnamefont {C.~L.}\ \bibnamefont
  {Kane}}\ and\ \bibinfo {author} {\bibfnamefont {E.~J.}\ \bibnamefont
  {Mele}},\ }\href {\doibase 10.1103/PhysRevLett.95.146802} {\bibfield
  {journal} {\bibinfo  {journal} {Phys. Rev. Lett.}\ }\textbf {\bibinfo
  {volume} {95}},\ \bibinfo {pages} {146802} (\bibinfo {year}
  {2005})}\BibitemShut {NoStop}%
\bibitem [{\citenamefont {Hasan}\ and\ \citenamefont
  {Kane}(2010)}]{Hasan:2010}%
  \BibitemOpen
  \bibfield  {author} {\bibinfo {author} {\bibfnamefont {M.~Z.}\ \bibnamefont
  {Hasan}}\ and\ \bibinfo {author} {\bibfnamefont {C.~L.}\ \bibnamefont
  {Kane}},\ }\href {http://link.aps.org/doi/10.1103/RevModPhys.82.3045}
  {\bibfield  {journal} {\bibinfo  {journal} {Reviews of Modern Physics}\
  }\textbf {\bibinfo {volume} {82}} (\bibinfo {year} {2010})}\BibitemShut
  {NoStop}%
\bibitem [{\citenamefont {Qi}\ and\ \citenamefont {Zhang}(2011)}]{Qi:2011}%
  \BibitemOpen
  \bibfield  {author} {\bibinfo {author} {\bibfnamefont {X.-L.}\ \bibnamefont
  {Qi}}\ and\ \bibinfo {author} {\bibfnamefont {S.-C.}\ \bibnamefont {Zhang}},\
  }\href {\doibase 10.1103/RevModPhys.83.1057} {\bibfield  {journal} {\bibinfo
  {journal} {Rev. Mod. Phys.}\ }\textbf {\bibinfo {volume} {83}},\ \bibinfo
  {pages} {1057} (\bibinfo {year} {2011})}\BibitemShut {NoStop}%
\bibitem [{\citenamefont {Zhang}\ \emph {et~al.}(2011)\citenamefont {Zhang},
  \citenamefont {Chang}, \citenamefont {Zhang}, \citenamefont {Wen},
  \citenamefont {Feng}, \citenamefont {Li}, \citenamefont {Liu}, \citenamefont
  {He}, \citenamefont {Wang}, \citenamefont {Chen}, \citenamefont {Xue},
  \citenamefont {Ma},\ and\ \citenamefont {Wang}}]{Zhang:2011b}%
  \BibitemOpen
  \bibfield  {author} {\bibinfo {author} {\bibfnamefont {J.}~\bibnamefont
  {Zhang}}, \bibinfo {author} {\bibfnamefont {C.-Z.}\ \bibnamefont {Chang}},
  \bibinfo {author} {\bibfnamefont {Z.}~\bibnamefont {Zhang}}, \bibinfo
  {author} {\bibfnamefont {J.}~\bibnamefont {Wen}}, \bibinfo {author}
  {\bibfnamefont {X.}~\bibnamefont {Feng}}, \bibinfo {author} {\bibfnamefont
  {K.}~\bibnamefont {Li}}, \bibinfo {author} {\bibfnamefont {M.}~\bibnamefont
  {Liu}}, \bibinfo {author} {\bibfnamefont {K.}~\bibnamefont {He}}, \bibinfo
  {author} {\bibfnamefont {L.}~\bibnamefont {Wang}}, \bibinfo {author}
  {\bibfnamefont {X.}~\bibnamefont {Chen}}, \bibinfo {author} {\bibfnamefont
  {Q.-K.}\ \bibnamefont {Xue}}, \bibinfo {author} {\bibfnamefont
  {X.}~\bibnamefont {Ma}}, \ and\ \bibinfo {author} {\bibfnamefont
  {Y.}~\bibnamefont {Wang}},\ }\href {\doibase 10.1038/ncomms1588} {\bibfield
  {journal} {\bibinfo  {journal} {Nat. Comm.}\ }\textbf {\bibinfo {volume} {2}}
  (\bibinfo {year} {2011}),\ 10.1038/ncomms1588}\BibitemShut {NoStop}%
\bibitem [{\citenamefont {Souma}\ \emph {et~al.}(2012)\citenamefont {Souma},
  \citenamefont {Eto}, \citenamefont {Nomura}, \citenamefont {Nakayama},
  \citenamefont {Sato}, \citenamefont {Takahashi}, \citenamefont {Segawa},\
  and\ \citenamefont {Ando}}]{Souma:2012}%
  \BibitemOpen
  \bibfield  {author} {\bibinfo {author} {\bibfnamefont {S.}~\bibnamefont
  {Souma}}, \bibinfo {author} {\bibfnamefont {K.}~\bibnamefont {Eto}}, \bibinfo
  {author} {\bibfnamefont {M.}~\bibnamefont {Nomura}}, \bibinfo {author}
  {\bibfnamefont {K.}~\bibnamefont {Nakayama}}, \bibinfo {author}
  {\bibfnamefont {T.}~\bibnamefont {Sato}}, \bibinfo {author} {\bibfnamefont
  {T.}~\bibnamefont {Takahashi}}, \bibinfo {author} {\bibfnamefont
  {K.}~\bibnamefont {Segawa}}, \ and\ \bibinfo {author} {\bibfnamefont
  {Y.}~\bibnamefont {Ando}},\ }\href
  {http://link.aps.org/doi/10.1103/PhysRevLett.108.116801} {\bibfield
  {journal} {\bibinfo  {journal} {Physical Review Letters}\ }\textbf {\bibinfo
  {volume} {108}},\ \bibinfo {pages} {116801} (\bibinfo {year}
  {2012})}\BibitemShut {NoStop}%
\bibitem [{\citenamefont {Arakane}\ \emph {et~al.}(2012)\citenamefont
  {Arakane}, \citenamefont {Sato}, \citenamefont {Souma}, \citenamefont
  {Kosaka}, \citenamefont {Nakayama}, \citenamefont {Komatsu}, \citenamefont
  {Takahashi}, \citenamefont {Ren}, \citenamefont {Segawa},\ and\ \citenamefont
  {Ando}}]{Arakane:2012}%
  \BibitemOpen
  \bibfield  {author} {\bibinfo {author} {\bibfnamefont {T.}~\bibnamefont
  {Arakane}}, \bibinfo {author} {\bibfnamefont {T.}~\bibnamefont {Sato}},
  \bibinfo {author} {\bibfnamefont {S.}~\bibnamefont {Souma}}, \bibinfo
  {author} {\bibfnamefont {K.}~\bibnamefont {Kosaka}}, \bibinfo {author}
  {\bibfnamefont {K.}~\bibnamefont {Nakayama}}, \bibinfo {author}
  {\bibfnamefont {M.}~\bibnamefont {Komatsu}}, \bibinfo {author} {\bibfnamefont
  {T.}~\bibnamefont {Takahashi}}, \bibinfo {author} {\bibfnamefont
  {Z.}~\bibnamefont {Ren}}, \bibinfo {author} {\bibfnamefont {K.}~\bibnamefont
  {Segawa}}, \ and\ \bibinfo {author} {\bibfnamefont {Y.}~\bibnamefont
  {Ando}},\ }\href {\doibase 10.1038/ncomms1639} {\bibfield  {journal}
  {\bibinfo  {journal} {Nat. Commun.}\ }\textbf {\bibinfo {volume} {3}}
  (\bibinfo {year} {2012}),\ 10.1038/ncomms1639}\BibitemShut {NoStop}%
\bibitem [{\citenamefont {Hsieh}\ \emph {et~al.}(2008)\citenamefont {Hsieh},
  \citenamefont {Qian}, \citenamefont {Wray}, \citenamefont {Xia},
  \citenamefont {Hor}, \citenamefont {Cava},\ and\ \citenamefont
  {Hasan}}]{Hsieh:2008}%
  \BibitemOpen
  \bibfield  {author} {\bibinfo {author} {\bibfnamefont {D.}~\bibnamefont
  {Hsieh}}, \bibinfo {author} {\bibfnamefont {D.}~\bibnamefont {Qian}},
  \bibinfo {author} {\bibfnamefont {L.}~\bibnamefont {Wray}}, \bibinfo {author}
  {\bibfnamefont {Y.}~\bibnamefont {Xia}}, \bibinfo {author} {\bibfnamefont
  {Y.~S.}\ \bibnamefont {Hor}}, \bibinfo {author} {\bibfnamefont {R.~J.}\
  \bibnamefont {Cava}}, \ and\ \bibinfo {author} {\bibfnamefont {M.~Z.}\
  \bibnamefont {Hasan}},\ }\href {http://dx.doi.org/10.1038/nature06843}
  {\bibfield  {journal} {\bibinfo  {journal} {Nature}\ }\textbf {\bibinfo
  {volume} {452}},\ \bibinfo {pages} {970} (\bibinfo {year}
  {2008})}\BibitemShut {NoStop}%
\bibitem [{\citenamefont {Hsieh}\ \emph
  {et~al.}(2009{\natexlab{a}})\citenamefont {Hsieh}, \citenamefont {Xia},
  \citenamefont {Wray}, \citenamefont {Qian}, \citenamefont {Pal},
  \citenamefont {Dil}, \citenamefont {Osterwalder}, \citenamefont {Meier},
  \citenamefont {Bihlmayer}, \citenamefont {Kane}, \citenamefont {Hor},
  \citenamefont {Cava},\ and\ \citenamefont {Hasan}}]{Hsieh:2009}%
  \BibitemOpen
  \bibfield  {author} {\bibinfo {author} {\bibfnamefont {D.}~\bibnamefont
  {Hsieh}}, \bibinfo {author} {\bibfnamefont {Y.}~\bibnamefont {Xia}}, \bibinfo
  {author} {\bibfnamefont {L.}~\bibnamefont {Wray}}, \bibinfo {author}
  {\bibfnamefont {D.}~\bibnamefont {Qian}}, \bibinfo {author} {\bibfnamefont
  {A.}~\bibnamefont {Pal}}, \bibinfo {author} {\bibfnamefont {J.~H.}\
  \bibnamefont {Dil}}, \bibinfo {author} {\bibfnamefont {J.}~\bibnamefont
  {Osterwalder}}, \bibinfo {author} {\bibfnamefont {F.}~\bibnamefont {Meier}},
  \bibinfo {author} {\bibfnamefont {G.}~\bibnamefont {Bihlmayer}}, \bibinfo
  {author} {\bibfnamefont {C.~L.}\ \bibnamefont {Kane}}, \bibinfo {author}
  {\bibfnamefont {Y.~S.}\ \bibnamefont {Hor}}, \bibinfo {author} {\bibfnamefont
  {R.~J.}\ \bibnamefont {Cava}}, \ and\ \bibinfo {author} {\bibfnamefont
  {M.~Z.}\ \bibnamefont {Hasan}},\ }\href {\doibase 10.1126/science.1167733}
  {\bibfield  {journal} {\bibinfo  {journal} {Science}\ }\textbf {\bibinfo
  {volume} {323}},\ \bibinfo {pages} {919} (\bibinfo {year}
  {2009}{\natexlab{a}})},\ \Eprint
  {http://arxiv.org/abs/http://www.sciencemag.org/cgi/reprint/323/5916/919.pdf}
  {http://www.sciencemag.org/cgi/reprint/323/5916/919.pdf} \BibitemShut
  {NoStop}%
\bibitem [{\citenamefont {Xia}\ \emph {et~al.}(2009)\citenamefont {Xia},
  \citenamefont {Qian}, \citenamefont {Hsieh}, \citenamefont {Wray},
  \citenamefont {Pal}, \citenamefont {Lin}, \citenamefont {Bansil},
  \citenamefont {Grauer}, \citenamefont {Hor}, \citenamefont {Cava},\ and\
  \citenamefont {Hasan}}]{Xia:2009}%
  \BibitemOpen
  \bibfield  {author} {\bibinfo {author} {\bibfnamefont {Y.}~\bibnamefont
  {Xia}}, \bibinfo {author} {\bibfnamefont {D.}~\bibnamefont {Qian}}, \bibinfo
  {author} {\bibfnamefont {D.}~\bibnamefont {Hsieh}}, \bibinfo {author}
  {\bibfnamefont {L.}~\bibnamefont {Wray}}, \bibinfo {author} {\bibfnamefont
  {A.}~\bibnamefont {Pal}}, \bibinfo {author} {\bibfnamefont {H.}~\bibnamefont
  {Lin}}, \bibinfo {author} {\bibfnamefont {A.}~\bibnamefont {Bansil}},
  \bibinfo {author} {\bibfnamefont {D.}~\bibnamefont {Grauer}}, \bibinfo
  {author} {\bibfnamefont {Y.~S.}\ \bibnamefont {Hor}}, \bibinfo {author}
  {\bibfnamefont {R.~J.}\ \bibnamefont {Cava}}, \ and\ \bibinfo {author}
  {\bibfnamefont {M.~Z.}\ \bibnamefont {Hasan}},\ }\href@noop {} {\bibfield
  {journal} {\bibinfo  {journal} {Nature Physics}\ }\textbf {\bibinfo {volume}
  {5}},\ \bibinfo {pages} {398} (\bibinfo {year} {2009})}\BibitemShut {NoStop}%
\bibitem [{\citenamefont {Chen}\ \emph {et~al.}(2009)\citenamefont {Chen},
  \citenamefont {Analytis}, \citenamefont {Chu}, \citenamefont {Liu},
  \citenamefont {Mo}, \citenamefont {Qi}, \citenamefont {Zhang}, \citenamefont
  {Lu}, \citenamefont {Dai}, \citenamefont {Fang}, \citenamefont {Zhang},
  \citenamefont {Fisher}, \citenamefont {Hussain},\ and\ \citenamefont
  {Shen}}]{Chen:2009}%
  \BibitemOpen
  \bibfield  {author} {\bibinfo {author} {\bibfnamefont {Y.~L.}\ \bibnamefont
  {Chen}}, \bibinfo {author} {\bibfnamefont {J.~G.}\ \bibnamefont {Analytis}},
  \bibinfo {author} {\bibfnamefont {J.-H.}\ \bibnamefont {Chu}}, \bibinfo
  {author} {\bibfnamefont {Z.~K.}\ \bibnamefont {Liu}}, \bibinfo {author}
  {\bibfnamefont {S.-K.}\ \bibnamefont {Mo}}, \bibinfo {author} {\bibfnamefont
  {X.~L.}\ \bibnamefont {Qi}}, \bibinfo {author} {\bibfnamefont {H.~J.}\
  \bibnamefont {Zhang}}, \bibinfo {author} {\bibfnamefont {D.~H.}\ \bibnamefont
  {Lu}}, \bibinfo {author} {\bibfnamefont {X.}~\bibnamefont {Dai}}, \bibinfo
  {author} {\bibfnamefont {Z.}~\bibnamefont {Fang}}, \bibinfo {author}
  {\bibfnamefont {S.~C.}\ \bibnamefont {Zhang}}, \bibinfo {author}
  {\bibfnamefont {I.~R.}\ \bibnamefont {Fisher}}, \bibinfo {author}
  {\bibfnamefont {Z.}~\bibnamefont {Hussain}}, \ and\ \bibinfo {author}
  {\bibfnamefont {Z.-X.}\ \bibnamefont {Shen}},\ }\href {\doibase
  10.1126/science.1173034} {\bibfield  {journal} {\bibinfo  {journal}
  {Science}\ }\textbf {\bibinfo {volume} {325}},\ \bibinfo {pages} {178}
  (\bibinfo {year} {2009})}\BibitemShut {NoStop}%
\bibitem [{\citenamefont {Nishide}\ \emph {et~al.}(2010)\citenamefont
  {Nishide}, \citenamefont {Taskin}, \citenamefont {Takeichi}, \citenamefont
  {Okuda}, \citenamefont {Kakizaki}, \citenamefont {Hirahara}, \citenamefont
  {Nakatsuji}, \citenamefont {Komori}, \citenamefont {Ando},\ and\
  \citenamefont {Matsuda}}]{Nishide:2010}%
  \BibitemOpen
  \bibfield  {author} {\bibinfo {author} {\bibfnamefont {A.}~\bibnamefont
  {Nishide}}, \bibinfo {author} {\bibfnamefont {A.~A.}\ \bibnamefont {Taskin}},
  \bibinfo {author} {\bibfnamefont {Y.}~\bibnamefont {Takeichi}}, \bibinfo
  {author} {\bibfnamefont {T.}~\bibnamefont {Okuda}}, \bibinfo {author}
  {\bibfnamefont {A.}~\bibnamefont {Kakizaki}}, \bibinfo {author}
  {\bibfnamefont {T.}~\bibnamefont {Hirahara}}, \bibinfo {author}
  {\bibfnamefont {K.}~\bibnamefont {Nakatsuji}}, \bibinfo {author}
  {\bibfnamefont {F.}~\bibnamefont {Komori}}, \bibinfo {author} {\bibfnamefont
  {Y.}~\bibnamefont {Ando}}, \ and\ \bibinfo {author} {\bibfnamefont
  {I.}~\bibnamefont {Matsuda}},\ }\href {\doibase 10.1103/PhysRevB.81.041309}
  {\bibfield  {journal} {\bibinfo  {journal} {Phys. Rev. B}\ }\textbf {\bibinfo
  {volume} {81}},\ \bibinfo {pages} {041309} (\bibinfo {year}
  {2010})}\BibitemShut {NoStop}%
\bibitem [{\citenamefont {Chen}\ \emph {et~al.}(2010)\citenamefont {Chen},
  \citenamefont {Liu}, \citenamefont {Analytis}, \citenamefont {Chu},
  \citenamefont {Zhang}, \citenamefont {Yan}, \citenamefont {Mo}, \citenamefont
  {Moore}, \citenamefont {Lu}, \citenamefont {Fisher}, \citenamefont {Zhang},
  \citenamefont {Hussain},\ and\ \citenamefont {Shen}}]{Chen:2010b}%
  \BibitemOpen
  \bibfield  {author} {\bibinfo {author} {\bibfnamefont {Y.~L.}\ \bibnamefont
  {Chen}}, \bibinfo {author} {\bibfnamefont {Z.~K.}\ \bibnamefont {Liu}},
  \bibinfo {author} {\bibfnamefont {J.~G.}\ \bibnamefont {Analytis}}, \bibinfo
  {author} {\bibfnamefont {J.-H.}\ \bibnamefont {Chu}}, \bibinfo {author}
  {\bibfnamefont {H.~J.}\ \bibnamefont {Zhang}}, \bibinfo {author}
  {\bibfnamefont {B.~H.}\ \bibnamefont {Yan}}, \bibinfo {author} {\bibfnamefont
  {S.-K.}\ \bibnamefont {Mo}}, \bibinfo {author} {\bibfnamefont {R.~G.}\
  \bibnamefont {Moore}}, \bibinfo {author} {\bibfnamefont {D.~H.}\ \bibnamefont
  {Lu}}, \bibinfo {author} {\bibfnamefont {I.~R.}\ \bibnamefont {Fisher}},
  \bibinfo {author} {\bibfnamefont {S.~C.}\ \bibnamefont {Zhang}}, \bibinfo
  {author} {\bibfnamefont {Z.}~\bibnamefont {Hussain}}, \ and\ \bibinfo
  {author} {\bibfnamefont {Z.-X.}\ \bibnamefont {Shen}},\ }\href {\doibase
  10.1103/PhysRevLett.105.266401} {\bibfield  {journal} {\bibinfo  {journal}
  {Phys. Rev. Lett.}\ }\textbf {\bibinfo {volume} {105}},\ \bibinfo {pages}
  {266401} (\bibinfo {year} {2010})}\BibitemShut {NoStop}%
\bibitem [{\citenamefont {Kuznetsova}\ \emph {et~al.}(2000)\citenamefont
  {Kuznetsova}, \citenamefont {Kuznetsov},\ and\ \citenamefont
  {Rowe}}]{Kuznetsova:2000}%
  \BibitemOpen
  \bibfield  {author} {\bibinfo {author} {\bibfnamefont {L.}~\bibnamefont
  {Kuznetsova}}, \bibinfo {author} {\bibfnamefont {V.}~\bibnamefont
  {Kuznetsov}}, \ and\ \bibinfo {author} {\bibfnamefont {D.}~\bibnamefont
  {Rowe}},\ }\href {\doibase 10.1016/S0022-3697(99)00423-0} {\bibfield
  {journal} {\bibinfo  {journal} {Journal of Physics and Chemistry of Solids}\
  }\textbf {\bibinfo {volume} {61}},\ \bibinfo {pages} {1269 } (\bibinfo {year}
  {2000})}\BibitemShut {NoStop}%
\bibitem [{\citenamefont {Zhang}\ \emph {et~al.}(2009)\citenamefont {Zhang},
  \citenamefont {Liu}, \citenamefont {Qi}, \citenamefont {Dai}, \citenamefont
  {Fang},\ and\ \citenamefont {Zhang}}]{Zhang:2009}%
  \BibitemOpen
  \bibfield  {author} {\bibinfo {author} {\bibfnamefont {H.}~\bibnamefont
  {Zhang}}, \bibinfo {author} {\bibfnamefont {C.-X.}\ \bibnamefont {Liu}},
  \bibinfo {author} {\bibfnamefont {X.-L.}\ \bibnamefont {Qi}}, \bibinfo
  {author} {\bibfnamefont {X.}~\bibnamefont {Dai}}, \bibinfo {author}
  {\bibfnamefont {Z.}~\bibnamefont {Fang}}, \ and\ \bibinfo {author}
  {\bibfnamefont {S.-C.}\ \bibnamefont {Zhang}},\ }\href@noop {} {\bibfield
  {journal} {\bibinfo  {journal} {Nature Physics}\ }\textbf {\bibinfo {volume}
  {5}},\ \bibinfo {pages} {438} (\bibinfo {year} {2009})}\BibitemShut {NoStop}%
\bibitem [{\citenamefont {Xu}\ \emph {et~al.}(2011)\citenamefont {Xu},
  \citenamefont {Wray}, \citenamefont {Xia}, \citenamefont {von Rohr},
  \citenamefont {Hor}, \citenamefont {Dil}, \citenamefont {Meier},
  \citenamefont {Slomski}, \citenamefont {Osterwalder}, \citenamefont
  {Neupane}, \citenamefont {Lin}, \citenamefont {Bansil}, \citenamefont
  {Fedorov}, \citenamefont {Cava},\ and\ \citenamefont {Hasan}}]{Xu:2011arXiv}%
  \BibitemOpen
  \bibfield  {author} {\bibinfo {author} {\bibfnamefont {S.-Y.}\ \bibnamefont
  {Xu}}, \bibinfo {author} {\bibfnamefont {L.~A.}\ \bibnamefont {Wray}},
  \bibinfo {author} {\bibfnamefont {Y.}~\bibnamefont {Xia}}, \bibinfo {author}
  {\bibfnamefont {F.}~\bibnamefont {von Rohr}}, \bibinfo {author}
  {\bibfnamefont {Y.~S.}\ \bibnamefont {Hor}}, \bibinfo {author} {\bibfnamefont
  {J.~H.}\ \bibnamefont {Dil}}, \bibinfo {author} {\bibfnamefont
  {F.}~\bibnamefont {Meier}}, \bibinfo {author} {\bibfnamefont
  {B.}~\bibnamefont {Slomski}}, \bibinfo {author} {\bibfnamefont
  {J.}~\bibnamefont {Osterwalder}}, \bibinfo {author} {\bibfnamefont
  {M.}~\bibnamefont {Neupane}}, \bibinfo {author} {\bibfnamefont
  {H.}~\bibnamefont {Lin}}, \bibinfo {author} {\bibfnamefont {A.}~\bibnamefont
  {Bansil}}, \bibinfo {author} {\bibfnamefont {A.}~\bibnamefont {Fedorov}},
  \bibinfo {author} {\bibfnamefont {R.~J.}\ \bibnamefont {Cava}}, \ and\
  \bibinfo {author} {\bibfnamefont {M.~Z.}\ \bibnamefont {Hasan}},\ }\href@noop
  {} {\bibfield  {journal} {\bibinfo  {journal} {ArXiv e-prints}\ } (\bibinfo
  {year} {2011})},\ \Eprint {http://arxiv.org/abs/1101.3985} {arXiv:1101.3985}
  \BibitemShut {NoStop}%
\bibitem [{\citenamefont {von Rohr}\ \emph {et~al.}(2013)\citenamefont {von
  Rohr}, \citenamefont {Schilling},\ and\ \citenamefont {Cava}}]{vRohr:2012}%
  \BibitemOpen
  \bibfield  {author} {\bibinfo {author} {\bibfnamefont {F.}~\bibnamefont {von
  Rohr}}, \bibinfo {author} {\bibfnamefont {A.}~\bibnamefont {Schilling}}, \
  and\ \bibinfo {author} {\bibfnamefont {R.~J.}\ \bibnamefont {Cava}},\ }\href
  {http://stacks.iop.org/0953-8984/25/i=7/a=075804} {\bibfield  {journal}
  {\bibinfo  {journal} {Journal of Physics: Condensed Matter}\ }\textbf
  {\bibinfo {volume} {25}},\ \bibinfo {pages} {075804} (\bibinfo {year}
  {2013})}\BibitemShut {NoStop}%
\bibitem [{\citenamefont {Hsieh}\ \emph
  {et~al.}(2009{\natexlab{b}})\citenamefont {Hsieh}, \citenamefont {Xia},
  \citenamefont {Qian}, \citenamefont {Wray}, \citenamefont {Dil},
  \citenamefont {Meier}, \citenamefont {Osterwalder}, \citenamefont {Patthey},
  \citenamefont {Checkelsky}, \citenamefont {Ong}, \citenamefont {Fedorov},
  \citenamefont {Lin}, \citenamefont {Bansil}, \citenamefont {Grauer},
  \citenamefont {Hor}, \citenamefont {Cava},\ and\ \citenamefont
  {Hasan}}]{Hsieh:2009N}%
  \BibitemOpen
  \bibfield  {author} {\bibinfo {author} {\bibfnamefont {D.}~\bibnamefont
  {Hsieh}}, \bibinfo {author} {\bibfnamefont {Y.}~\bibnamefont {Xia}}, \bibinfo
  {author} {\bibfnamefont {D.}~\bibnamefont {Qian}}, \bibinfo {author}
  {\bibfnamefont {L.}~\bibnamefont {Wray}}, \bibinfo {author} {\bibfnamefont
  {J.}~\bibnamefont {Dil}}, \bibinfo {author} {\bibfnamefont {F.}~\bibnamefont
  {Meier}}, \bibinfo {author} {\bibfnamefont {J.}~\bibnamefont {Osterwalder}},
  \bibinfo {author} {\bibfnamefont {L.}~\bibnamefont {Patthey}}, \bibinfo
  {author} {\bibfnamefont {J.}~\bibnamefont {Checkelsky}}, \bibinfo {author}
  {\bibfnamefont {N.}~\bibnamefont {Ong}}, \bibinfo {author} {\bibfnamefont
  {A.}~\bibnamefont {Fedorov}}, \bibinfo {author} {\bibfnamefont
  {H.}~\bibnamefont {Lin}}, \bibinfo {author} {\bibfnamefont {A.}~\bibnamefont
  {Bansil}}, \bibinfo {author} {\bibfnamefont {D.}~\bibnamefont {Grauer}},
  \bibinfo {author} {\bibfnamefont {Y.}~\bibnamefont {Hor}}, \bibinfo {author}
  {\bibfnamefont {R.}~\bibnamefont {Cava}}, \ and\ \bibinfo {author}
  {\bibfnamefont {M.}~\bibnamefont {Hasan}},\ }\href@noop {} {\bibfield
  {journal} {\bibinfo  {journal} {Nature}\ }\textbf {\bibinfo {volume} {460}},\
  \bibinfo {pages} {1101} (\bibinfo {year} {2009}{\natexlab{b}})}\BibitemShut
  {NoStop}%
\bibitem [{\citenamefont {Hoesch}\ \emph {et~al.}(2002)\citenamefont {Hoesch},
  \citenamefont {Greber}, \citenamefont {Petrov}, \citenamefont {Muntwiler},
  \citenamefont {Hengsberger}, \citenamefont {Auwaerter},\ and\ \citenamefont
  {Osterwalder}}]{Hoesch:2002}%
  \BibitemOpen
  \bibfield  {author} {\bibinfo {author} {\bibfnamefont {M.}~\bibnamefont
  {Hoesch}}, \bibinfo {author} {\bibfnamefont {T.}~\bibnamefont {Greber}},
  \bibinfo {author} {\bibfnamefont {V.~N.}\ \bibnamefont {Petrov}}, \bibinfo
  {author} {\bibfnamefont {M.}~\bibnamefont {Muntwiler}}, \bibinfo {author}
  {\bibfnamefont {M.}~\bibnamefont {Hengsberger}}, \bibinfo {author}
  {\bibfnamefont {W.}~\bibnamefont {Auwaerter}}, \ and\ \bibinfo {author}
  {\bibfnamefont {J.}~\bibnamefont {Osterwalder}},\ }\href {\doibase DOI:
  10.1016/S0368-2048(02)00058-0} {\bibfield  {journal} {\bibinfo  {journal}
  {Journal of Electron Spectroscopy and Related Phenomena}\ }\textbf {\bibinfo
  {volume} {124}},\ \bibinfo {pages} {263 } (\bibinfo {year}
  {2002})}\BibitemShut {NoStop}%
\bibitem [{\citenamefont {Eremeev}\ \emph {et~al.}(2012)\citenamefont
  {Eremeev}, \citenamefont {Landolt}, \citenamefont {Menshchikova},
  \citenamefont {Slomski}, \citenamefont {Koroteev}, \citenamefont {Aliev},
  \citenamefont {Babanly}, \citenamefont {Henk}, \citenamefont {Ernst},
  \citenamefont {Patthey}, \citenamefont {Eich}, \citenamefont {Khajetoorians},
  \citenamefont {Hagemeister}, \citenamefont {Pietzsch}, \citenamefont {Wiebe},
  \citenamefont {Wiesendanger}, \citenamefont {Echenique}, \citenamefont
  {Tsirkin}, \citenamefont {Amiraslanov}, \citenamefont {Dil},\ and\
  \citenamefont {Chulkov}}]{Eremeev:2012}%
  \BibitemOpen
  \bibfield  {author} {\bibinfo {author} {\bibfnamefont {S.~V.}\ \bibnamefont
  {Eremeev}}, \bibinfo {author} {\bibfnamefont {G.}~\bibnamefont {Landolt}},
  \bibinfo {author} {\bibfnamefont {T.~V.}\ \bibnamefont {Menshchikova}},
  \bibinfo {author} {\bibfnamefont {B.}~\bibnamefont {Slomski}}, \bibinfo
  {author} {\bibfnamefont {Y.~M.}\ \bibnamefont {Koroteev}}, \bibinfo {author}
  {\bibfnamefont {Z.~S.}\ \bibnamefont {Aliev}}, \bibinfo {author}
  {\bibfnamefont {M.~B.}\ \bibnamefont {Babanly}}, \bibinfo {author}
  {\bibfnamefont {J.}~\bibnamefont {Henk}}, \bibinfo {author} {\bibfnamefont
  {A.}~\bibnamefont {Ernst}}, \bibinfo {author} {\bibfnamefont
  {L.}~\bibnamefont {Patthey}}, \bibinfo {author} {\bibfnamefont
  {A.}~\bibnamefont {Eich}}, \bibinfo {author} {\bibfnamefont {A.~A.}\
  \bibnamefont {Khajetoorians}}, \bibinfo {author} {\bibfnamefont
  {J.}~\bibnamefont {Hagemeister}}, \bibinfo {author} {\bibfnamefont
  {O.}~\bibnamefont {Pietzsch}}, \bibinfo {author} {\bibfnamefont
  {J.}~\bibnamefont {Wiebe}}, \bibinfo {author} {\bibfnamefont
  {R.}~\bibnamefont {Wiesendanger}}, \bibinfo {author} {\bibfnamefont {P.~M.}\
  \bibnamefont {Echenique}}, \bibinfo {author} {\bibfnamefont {S.~S.}\
  \bibnamefont {Tsirkin}}, \bibinfo {author} {\bibfnamefont {I.~R.}\
  \bibnamefont {Amiraslanov}}, \bibinfo {author} {\bibfnamefont {J.~H.}\
  \bibnamefont {Dil}}, \ and\ \bibinfo {author} {\bibfnamefont {E.~V.}\
  \bibnamefont {Chulkov}},\ }\href {http://dx.doi.org/10.1038/ncomms1638}
  {\bibfield  {journal} {\bibinfo  {journal} {Nat Commun}\ }\textbf {\bibinfo
  {volume} {3}},\ \bibinfo {pages} {103} (\bibinfo {year} {2012})}\BibitemShut
  {NoStop}%
\bibitem [{\citenamefont {Shelimova}\ \emph {et~al.}(2004)\citenamefont
  {Shelimova}, \citenamefont {Karpinskii}, \citenamefont {Konstantinov},
  \citenamefont {Avilov}, \citenamefont {Kretova},\ and\ \citenamefont
  {Zemskov}}]{Shelimova:2004}%
  \BibitemOpen
  \bibfield  {author} {\bibinfo {author} {\bibfnamefont {L.}~\bibnamefont
  {Shelimova}}, \bibinfo {author} {\bibfnamefont {O.}~\bibnamefont
  {Karpinskii}}, \bibinfo {author} {\bibfnamefont {P.}~\bibnamefont
  {Konstantinov}}, \bibinfo {author} {\bibfnamefont {E.}~\bibnamefont
  {Avilov}}, \bibinfo {author} {\bibfnamefont {M.}~\bibnamefont {Kretova}}, \
  and\ \bibinfo {author} {\bibfnamefont {V.}~\bibnamefont {Zemskov}},\
  }\href@noop {} {\bibfield  {journal} {\bibinfo  {journal} {Inorganic
  Materials}\ }\textbf {\bibinfo {volume} {40}},\ \bibinfo {pages} {451}
  (\bibinfo {year} {2004})}\BibitemShut {NoStop}%
\bibitem [{\citenamefont {Kuznetsov}\ \emph {et~al.}(1998)\citenamefont
  {Kuznetsov}, \citenamefont {Kuznetsova},\ and\ \citenamefont
  {Rowe}}]{Kuznetsov:1998}%
  \BibitemOpen
  \bibfield  {author} {\bibinfo {author} {\bibfnamefont {V.~L.}\ \bibnamefont
  {Kuznetsov}}, \bibinfo {author} {\bibfnamefont {L.~A.}\ \bibnamefont
  {Kuznetsova}}, \ and\ \bibinfo {author} {\bibfnamefont {D.~M.}\ \bibnamefont
  {Rowe}},\ }\href {http://dx.doi.org/10.1063/1.369662} {\bibfield  {journal}
  {\bibinfo  {journal} {Journal of Applied Physics}\ }\textbf {\bibinfo
  {volume} {85}},\ \bibinfo {pages} {3207} (\bibinfo {year}
  {1998})}\BibitemShut {NoStop}%
\bibitem [{\citenamefont {Okamoto}\ \emph {et~al.}(2012)\citenamefont
  {Okamoto}, \citenamefont {Kuroda}, \citenamefont {Miyahara}, \citenamefont
  {Miyamoto}, \citenamefont {Okuda}, \citenamefont {Aliev}, \citenamefont
  {Babanly}, \citenamefont {Amiraslanov}, \citenamefont {Shimada},
  \citenamefont {Namatame}, \citenamefont {Taniguchi}, \citenamefont
  {Samorokov}, \citenamefont {Menshchikova}, \citenamefont {Chulkov},\ and\
  \citenamefont {Kimura}}]{Okamoto:2012}%
  \BibitemOpen
  \bibfield  {author} {\bibinfo {author} {\bibfnamefont {K.}~\bibnamefont
  {Okamoto}}, \bibinfo {author} {\bibfnamefont {K.}~\bibnamefont {Kuroda}},
  \bibinfo {author} {\bibfnamefont {H.}~\bibnamefont {Miyahara}}, \bibinfo
  {author} {\bibfnamefont {K.}~\bibnamefont {Miyamoto}}, \bibinfo {author}
  {\bibfnamefont {T.}~\bibnamefont {Okuda}}, \bibinfo {author} {\bibfnamefont
  {Z.~S.}\ \bibnamefont {Aliev}}, \bibinfo {author} {\bibfnamefont {M.~B.}\
  \bibnamefont {Babanly}}, \bibinfo {author} {\bibfnamefont {I.~R.}\
  \bibnamefont {Amiraslanov}}, \bibinfo {author} {\bibfnamefont
  {K.}~\bibnamefont {Shimada}}, \bibinfo {author} {\bibfnamefont
  {H.}~\bibnamefont {Namatame}}, \bibinfo {author} {\bibfnamefont
  {M.}~\bibnamefont {Taniguchi}}, \bibinfo {author} {\bibfnamefont {D.~A.}\
  \bibnamefont {Samorokov}}, \bibinfo {author} {\bibfnamefont {T.~V.}\
  \bibnamefont {Menshchikova}}, \bibinfo {author} {\bibfnamefont {E.~V.}\
  \bibnamefont {Chulkov}}, \ and\ \bibinfo {author} {\bibfnamefont
  {A.}~\bibnamefont {Kimura}},\ }\href {\doibase 10.1103/PhysRevB.86.195304}
  {\bibfield  {journal} {\bibinfo  {journal} {Phys. Rev. B}\ }\textbf {\bibinfo
  {volume} {86}},\ \bibinfo {pages} {195304} (\bibinfo {year}
  {2012})}\BibitemShut {NoStop}%
\bibitem [{\citenamefont {Park}\ and\ \citenamefont
  {Louie}(2012)}]{Park:2012a}%
  \BibitemOpen
  \bibfield  {author} {\bibinfo {author} {\bibfnamefont {C.-H.}\ \bibnamefont
  {Park}}\ and\ \bibinfo {author} {\bibfnamefont {S.~G.}\ \bibnamefont
  {Louie}},\ }\href {\doibase 10.1103/PhysRevLett.109.097601} {\bibfield
  {journal} {\bibinfo  {journal} {Phys. Rev. Lett.}\ }\textbf {\bibinfo
  {volume} {109}},\ \bibinfo {pages} {097601} (\bibinfo {year}
  {2012})}\BibitemShut {NoStop}%
\bibitem [{\citenamefont {Fu}(2009)}]{Fu:2009w}%
  \BibitemOpen
  \bibfield  {author} {\bibinfo {author} {\bibfnamefont {L.}~\bibnamefont
  {Fu}},\ }\href {\doibase 10.1103/PhysRevLett.103.266801} {\bibfield
  {journal} {\bibinfo  {journal} {Phys. Rev. Lett.}\ }\textbf {\bibinfo
  {volume} {103}},\ \bibinfo {pages} {266801} (\bibinfo {year}
  {2009})}\BibitemShut {NoStop}%
\bibitem [{\citenamefont {Souma}\ \emph {et~al.}(2011)\citenamefont {Souma},
  \citenamefont {Kosaka}, \citenamefont {Sato}, \citenamefont {Komatsu},
  \citenamefont {Takayama}, \citenamefont {Takahashi}, \citenamefont {Kriener},
  \citenamefont {Segawa},\ and\ \citenamefont {Ando}}]{Souma:2011}%
  \BibitemOpen
  \bibfield  {author} {\bibinfo {author} {\bibfnamefont {S.}~\bibnamefont
  {Souma}}, \bibinfo {author} {\bibfnamefont {K.}~\bibnamefont {Kosaka}},
  \bibinfo {author} {\bibfnamefont {T.}~\bibnamefont {Sato}}, \bibinfo {author}
  {\bibfnamefont {M.}~\bibnamefont {Komatsu}}, \bibinfo {author} {\bibfnamefont
  {A.}~\bibnamefont {Takayama}}, \bibinfo {author} {\bibfnamefont
  {T.}~\bibnamefont {Takahashi}}, \bibinfo {author} {\bibfnamefont
  {M.}~\bibnamefont {Kriener}}, \bibinfo {author} {\bibfnamefont
  {K.}~\bibnamefont {Segawa}}, \ and\ \bibinfo {author} {\bibfnamefont
  {Y.}~\bibnamefont {Ando}},\ }\href {\doibase 10.1103/PhysRevLett.106.216803}
  {\bibfield  {journal} {\bibinfo  {journal} {Phys. Rev. Lett.}\ }\textbf
  {\bibinfo {volume} {106}},\ \bibinfo {pages} {216803} (\bibinfo {year}
  {2011})}\BibitemShut {NoStop}%
\bibitem [{\citenamefont {Dil}(2009)}]{Dil:2009R}%
  \BibitemOpen
  \bibfield  {author} {\bibinfo {author} {\bibfnamefont {J.~H.}\ \bibnamefont
  {Dil}},\ }\href {http://stacks.iop.org/0953-8984/21/403001} {\bibfield
  {journal} {\bibinfo  {journal} {Journal of Physics: Condensed Matter}\
  }\textbf {\bibinfo {volume} {21}},\ \bibinfo {pages} {403001 (22pp)}
  (\bibinfo {year} {2009})}\BibitemShut {NoStop}%
\bibitem [{\citenamefont {Zhang}\ and\ \citenamefont
  {Yates}(2012)}]{Zhang:2012}%
  \BibitemOpen
  \bibfield  {author} {\bibinfo {author} {\bibfnamefont {Z.}~\bibnamefont
  {Zhang}}\ and\ \bibinfo {author} {\bibfnamefont {J.~T.}\ \bibnamefont
  {Yates}},\ }\href {\doibase 10.1021/cr3000626} {\bibfield  {journal}
  {\bibinfo  {journal} {Chemical Reviews}\ }\textbf {\bibinfo {volume} {112}},\
  \bibinfo {pages} {5520} (\bibinfo {year} {2012})}\BibitemShut {NoStop}%
\bibitem [{\citenamefont {Bianchi}\ \emph {et~al.}(2010)\citenamefont
  {Bianchi}, \citenamefont {Guan}, \citenamefont {Bao}, \citenamefont {Mi},
  \citenamefont {Iversen}, \citenamefont {King},\ and\ \citenamefont
  {Hofmann}}]{Bianchi:2010}%
  \BibitemOpen
  \bibfield  {author} {\bibinfo {author} {\bibfnamefont {M.}~\bibnamefont
  {Bianchi}}, \bibinfo {author} {\bibfnamefont {D.}~\bibnamefont {Guan}},
  \bibinfo {author} {\bibfnamefont {S.}~\bibnamefont {Bao}}, \bibinfo {author}
  {\bibfnamefont {J.}~\bibnamefont {Mi}}, \bibinfo {author} {\bibfnamefont
  {B.~B.}\ \bibnamefont {Iversen}}, \bibinfo {author} {\bibfnamefont
  {P.~D.~C.}\ \bibnamefont {King}}, \ and\ \bibinfo {author} {\bibfnamefont
  {P.}~\bibnamefont {Hofmann}},\ }\href {http://dx.doi.org/10.1038/ncomms1131}
  {\bibfield  {journal} {\bibinfo  {journal} {Nat Commun}\ }\textbf {\bibinfo
  {volume} {1}} (\bibinfo {year} {2010})}\BibitemShut {NoStop}%
\bibitem [{\citenamefont {King}\ \emph {et~al.}(2011)\citenamefont {King},
  \citenamefont {Hatch}, \citenamefont {Bianchi}, \citenamefont {Ovsyannikov},
  \citenamefont {Lupulescu}, \citenamefont {Landolt}, \citenamefont {Slomski},
  \citenamefont {Dil}, \citenamefont {Guan}, \citenamefont {Mi}, \citenamefont
  {Rienks}, \citenamefont {Fink}, \citenamefont {Lindblad}, \citenamefont
  {Svensson}, \citenamefont {Bao}, \citenamefont {Balakrishnan}, \citenamefont
  {Iversen}, \citenamefont {Osterwalder}, \citenamefont {Eberhardt},
  \citenamefont {Baumberger},\ and\ \citenamefont {Hofmann}}]{King:2011}%
  \BibitemOpen
  \bibfield  {author} {\bibinfo {author} {\bibfnamefont {P.~D.~C.}\
  \bibnamefont {King}}, \bibinfo {author} {\bibfnamefont {R.~C.}\ \bibnamefont
  {Hatch}}, \bibinfo {author} {\bibfnamefont {M.}~\bibnamefont {Bianchi}},
  \bibinfo {author} {\bibfnamefont {R.}~\bibnamefont {Ovsyannikov}}, \bibinfo
  {author} {\bibfnamefont {C.}~\bibnamefont {Lupulescu}}, \bibinfo {author}
  {\bibfnamefont {G.}~\bibnamefont {Landolt}}, \bibinfo {author} {\bibfnamefont
  {B.}~\bibnamefont {Slomski}}, \bibinfo {author} {\bibfnamefont {J.~H.}\
  \bibnamefont {Dil}}, \bibinfo {author} {\bibfnamefont {D.}~\bibnamefont
  {Guan}}, \bibinfo {author} {\bibfnamefont {J.~L.}\ \bibnamefont {Mi}},
  \bibinfo {author} {\bibfnamefont {E.~D.~L.}\ \bibnamefont {Rienks}}, \bibinfo
  {author} {\bibfnamefont {J.}~\bibnamefont {Fink}}, \bibinfo {author}
  {\bibfnamefont {A.}~\bibnamefont {Lindblad}}, \bibinfo {author}
  {\bibfnamefont {S.}~\bibnamefont {Svensson}}, \bibinfo {author}
  {\bibfnamefont {S.}~\bibnamefont {Bao}}, \bibinfo {author} {\bibfnamefont
  {G.}~\bibnamefont {Balakrishnan}}, \bibinfo {author} {\bibfnamefont {B.~B.}\
  \bibnamefont {Iversen}}, \bibinfo {author} {\bibfnamefont {J.}~\bibnamefont
  {Osterwalder}}, \bibinfo {author} {\bibfnamefont {W.}~\bibnamefont
  {Eberhardt}}, \bibinfo {author} {\bibfnamefont {F.}~\bibnamefont
  {Baumberger}}, \ and\ \bibinfo {author} {\bibfnamefont {P.}~\bibnamefont
  {Hofmann}},\ }\href {http://link.aps.org/doi/10.1103/PhysRevLett.107.096802}
  {\bibfield  {journal} {\bibinfo  {journal} {Physical Review Letters}\
  }\textbf {\bibinfo {volume} {107}},\ \bibinfo {pages} {096802} (\bibinfo
  {year} {2011})}\BibitemShut {NoStop}%
\bibitem [{\citenamefont {Niu}\ \emph {et~al.}(2012)\citenamefont {Niu},
  \citenamefont {Dai}, \citenamefont {Zhu}, \citenamefont {Ma}, \citenamefont
  {Yu}, \citenamefont {Han},\ and\ \citenamefont {Huang}}]{Niu:2012}%
  \BibitemOpen
  \bibfield  {author} {\bibinfo {author} {\bibfnamefont {C.}~\bibnamefont
  {Niu}}, \bibinfo {author} {\bibfnamefont {Y.}~\bibnamefont {Dai}}, \bibinfo
  {author} {\bibfnamefont {Y.}~\bibnamefont {Zhu}}, \bibinfo {author}
  {\bibfnamefont {Y.}~\bibnamefont {Ma}}, \bibinfo {author} {\bibfnamefont
  {L.}~\bibnamefont {Yu}}, \bibinfo {author} {\bibfnamefont {S.}~\bibnamefont
  {Han}}, \ and\ \bibinfo {author} {\bibfnamefont {B.}~\bibnamefont {Huang}},\
  }\href {\doibase 10.1038/srep00976} {\bibfield  {journal} {\bibinfo
  {journal} {Sci. Rep.}\ }\textbf {\bibinfo {volume} {2}} (\bibinfo {year}
  {2012}),\ 10.1038/srep00976}\BibitemShut {NoStop}%
\bibitem [{\citenamefont {Huang}\ \emph {et~al.}(2012)\citenamefont {Huang},
  \citenamefont {Chu}, \citenamefont {Kung}, \citenamefont {Lee}, \citenamefont
  {Sankar}, \citenamefont {Liou}, \citenamefont {Wu}, \citenamefont {Kuo},\
  and\ \citenamefont {Chou}}]{Huang:2012b}%
  \BibitemOpen
  \bibfield  {author} {\bibinfo {author} {\bibfnamefont {F.-T.}\ \bibnamefont
  {Huang}}, \bibinfo {author} {\bibfnamefont {M.-W.}\ \bibnamefont {Chu}},
  \bibinfo {author} {\bibfnamefont {H.~H.}\ \bibnamefont {Kung}}, \bibinfo
  {author} {\bibfnamefont {W.~L.}\ \bibnamefont {Lee}}, \bibinfo {author}
  {\bibfnamefont {R.}~\bibnamefont {Sankar}}, \bibinfo {author} {\bibfnamefont
  {S.-C.}\ \bibnamefont {Liou}}, \bibinfo {author} {\bibfnamefont {K.~K.}\
  \bibnamefont {Wu}}, \bibinfo {author} {\bibfnamefont {Y.~K.}\ \bibnamefont
  {Kuo}}, \ and\ \bibinfo {author} {\bibfnamefont {F.~C.}\ \bibnamefont
  {Chou}},\ }\href {\doibase 10.1103/PhysRevB.86.081104} {\bibfield  {journal}
  {\bibinfo  {journal} {Phys. Rev. B}\ }\textbf {\bibinfo {volume} {86}},\
  \bibinfo {pages} {081104} (\bibinfo {year} {2012})}\BibitemShut {NoStop}%
\end{thebibliography}

%

\end{document}